\documentclass[aps,prb,twocolumn]{revtex4}%
\usepackage{amsmath}
\usepackage{amssymb}
\usepackage{graphicx}
\usepackage{amsfonts}%
\begin{document}
\title{Commensurability oscillations in the SAW induced acousto-electric effect in a 2DEG.}
\author{John P. Robinson and Vladimir I. Fal'ko}
\affiliation{Physics Department, Lancaster University, LA1 4YB, UK}
\date{\today}

\begin{abstract}
We study the acousto-electric (AE)\ effect generated by surface
acoustic waves (SAW) in a high mobility 2D electron gas (2DEG)
with isotropic and especially small-angle impurity scattering. In
both cases the acousto-electric effect exhibits Weiss oscillations
periodic in $B^{-1}$ due to the commensurability of the SAW period
with the size of the cyclotron orbit and resonances at the SAW
frequency $\omega=k\omega_{c}$ multiple of the cyclotron
frequency. We describe how oscillations in the acousto-electric
effect are damped in low fields where
$\omega_{c}\tau_{\ast}\lesssim1$ (with the time scale
$\tau_{\ast}$ dependent on the type of scattering) and find its
non-oscillatory part which remains finite to the lowest fields.
\end{abstract}

\maketitle

Due to a finite wave number carried by surface acoustic waves
(SAWs) their application enables one to access the properties of
low-dimensional electron systems\cite{Willett2} that cannot be
studied using the standard microwave absorption techniques. The
observation of magneto-oscillations in the shift of SAW velocity
caused by its interaction with the 2D electrons in the vicinity of
filling factor $\nu=1/2$ and its comparison with Weiss geometrical
oscillations\cite{Weiss,Klitzing,Beenakker} for electrons has
enabled Willett \textit{et al.}\cite{Willett} to establish the
existence of a composite fermion Fermi surface. Also, the
additional length scale in the system permits transitions
otherwise forbidden by Kohn's theorem\cite{Kohn}, thus making
possible detection of cyclotron transitions in a gas of composite
quasiparticles\cite{Simon,Kukushkin}.

In this Communication we extend the analysis of the phenomenon of
geometrical commensurability onto the acousto-electric (drag)
effect\cite{Efros,Bergli1,Bergli2,Falko1} which has been studied
experimentally in semiconductor structures by Esslinger \textit{et
al}.\cite{Esslingen} and Shilton \textit{et al.}\cite{Shilton}. We
show that by measuring magneto-oscillations and study the
frequency dependence of the DC electric field induced in a 2DEG by
a propagating SAW, one can access the same information about the
Fermi gas as was previously studied in absorption and SAW
propagation experiments\cite{Willett2,Kukushkin}. In the present
publication we also compare and contrast two types of
high-mobility structures: with isotropic and small-angle impurity
scattering.

Here we use an approach recently applied to the studies of another
DC effect produced by a dynamical acoustic wave field, the
SAW-induced magneto-resistance\cite{Robinson}. We investigate the
AE effect in the linear order in the SAW power and find the
parametric dependences of the steady-state electric field
$E_{\mathrm{AE}}$ generated by
the SAW in the direction of its propagation, $\mathbf{E_{\mathrm{AE}}%
}=\mathbf{\hat{q}}E_{\mathrm{AE}}\left(
qR_{c},\omega_{c}\tau_{\ast},\omega/\omega _{c}\right)  $ , where
$\mathbf{q}$ is the SAW wave vector, $R_{c}$ and $\omega_{c}$ the
electron cyclotron radius and frequency, $\omega$ the SAW
frequency, and $\tau_{\ast}$ is the effective electron scattering
time crucially dependent on the type of impurity
scattering\cite{Aleiner1,Mirlin1}. In the high-field limit,
$\omega_{c}\tau_{\ast}\gg1$, we describe Weiss oscillations of the
AE effect as a function of $qR_{c}$. In addition, for low-density
structures (as well as for heavy-mass carriers, such as composite
fermions) we predict resonances in $E_{\mathrm{AE}}$ at
frequencies $\omega=k\omega_{c}$. For a small magnetic field
(although large enough to ensure that $v_{F}B>E_{\omega q}$), we
show that while commensurability oscillations are damped, there is
a finite field-independent contribution to the AE drag.

Our theory consists of the analysis of the Boltzmann equation,
\begin{gather}
\mathcal{\hat{L}}\left[  f(\mathbf{p},\mathbf{x},t)\right]=\mathcal{\hat{C}%
}\left[  f(\mathbf{p},\mathbf{x},t)\right]\,,\label{Kinetic-Eq1}%
\\
\mathcal{\hat{L}}=\partial_{t}+\omega_{c}R_{c}\cos\varphi\partial_{x}%
+\omega_{c}\partial_{\varphi}+eE{\mathcal{\hat{P}}}\,,\nonumber \\
\mathcal{\hat{P}}=v\cos\varphi\partial_{\epsilon}-\frac{\sin\varphi}%
{p}\partial_{\varphi}\,,\nonumber
\end{gather}
where $\mathcal{\hat{C}}$ is the collision integral and the
momentum dependent part in $\mathcal{\hat{L}}$ and
$\mathcal{\hat{P}}$ is written in terms of the electron kinetic
energy $\epsilon=p^{2}/2m$ and angle $\varphi$, characterizing the
direction of electron propagation with respect to the direction of
propagation of the SAW. Here, $v$ is the electron velocity,
$\omega_{c}$ and $R_{c}$ are the cyclotron frequency and radius
respectively, $\mathbf{E}=\mathbf{l}_{x}Ee^{i\left(\omega
t-qx\right)}$ is the longitudinally polarized SAW field, screened
by the 2DEG.

Using the $(\epsilon,\varphi)$ parametrization of momentum space, we expand
the distribution function $f(\mathbf{p},\mathbf{x},t)$ into
\begin{equation}
f(\epsilon,\varphi,x,t)=f_{T}(\epsilon)+\sum_{m}\sum_{\Omega
Q}f_{\Omega
Q}^{m}(\epsilon)e^{-i\Omega t+iQx}e^{im\varphi}\,,\label{harmonics-of-f}%
\end{equation}
where $f_{T}(\epsilon)$ is the equilibrium Fermi function and
$f_{\omega q}^{m}$ characterize the non-equilibrium state caused
by the SAW. We determine $g^{m}=\int_{0}^{\infty}d\epsilon f^{m}$,
so that $g^{0}$ would characterize the electron density and
$g^{\pm1}$ combine into electric current. We separate the
collision integral
\begin{equation}
\mathcal{\hat{C}}\left[  f\left(  \epsilon,\varphi\right)  \right]
=\mathcal{\hat{C}}_{\sigma/\eta}\left[  f\left(  \epsilon,\varphi\right)
\right]  -\frac{f^{0}(\epsilon)+(\partial_{\epsilon}f_{T})g^{0}}{\tau_{in}%
},\label{Coll-int}%
\end{equation}
into the elastic and inelastic parts. Relaxation of the non-equilibrium part
of the distribution function towards an isotropic distribution is described by
the term $\mathcal{\hat{C}}_{\sigma/\eta}$,
\begin{equation}
\mathcal{\hat{C}}_{\eta}\left[  f\right]  =\frac{f^{0}-f}{\tau},\qquad
\mathrm{and}\qquad\mathcal{\hat{C}}_{\sigma}\left[  f\right]  =\frac{1}{\tau
}{\partial_{\varphi}^{2}}f.\label{Coll-int-long}%
\end{equation}
We adopt the subscripts $\eta$ for isotropic scattering and
$\sigma$ for small angle scattering, approximated by diffusion
along the Fermi surface. In Eq.~.(\ref{Coll-int-long}),
$\tau^{-1}$ is the momentum relaxation rate [the timescale upon
which the non-equilibrium harmonics $f^{\pm1}$ decay]. Energy
relaxation, with $\tau_{in}^{-1}\ll\tau^{-1}$, is taken into
account by the last term in Eq.~(\ref{Coll-int}) using the
relaxation time approximation.

The rectified (acousto-electric) current can be described using
\begin{equation}
\mathcal{J}=j_{x}-ij_{y}=e\gamma\int_{0}^{\infty}d\epsilon vf_{00}^{1}\left(
\epsilon\right)  ,
\end{equation}
where $\gamma$ is the 2D density of states, and $f_{00}\left(
\epsilon ,\varphi\right)  $ is the steady state homogenous part of
the non-equilibrium distribution. Below, we restrict the analysis
to effects linear in the SAW power and perform a perturbative
analysis. We assume that the force from the SAW field is much less
than the Lorentz force, $E_{\omega q}\ll v_{F}B$, whereby electron
cyclotron orbits are not destroyed by the SAW and no channelling
of electron trajectories occurs. To describe the AE\ effect we
relate the steady state term $f_{00}\left( \epsilon,\varphi\right)
$ to the SAW\ field and $f_{\omega q}\left(
\epsilon,\varphi\right)  $ at the SAW frequency by taking the
$Q=0$, $\Omega=0$ harmonics of Eq.~(\ref{Kinetic-Eq1}),
\begin{equation}
\partial_{\varphi}f_{00}(\epsilon,\varphi)=\frac{\mathcal{\hat{C}}%
f_{00}\left(  \epsilon,\varphi\right)  }{\omega_{c}}-\sum_{\pm\omega q}%
\frac{eE_{-\omega-q}}{\omega_{c}}{\mathcal{\hat{P}}}f_{\omega q}%
(\epsilon,\varphi).\nonumber
\end{equation}
We then evaluate the complex current, $\mathcal{J}$,
\begin{eqnarray}
&\mathcal{J}&=-\sum_{\pm\omega
q}\frac{e^{2}E_{-\omega-q}}{\omega_{c}}\int
_{0}^{\infty}d\epsilon\frac{\gamma\omega_{c}\tau}{\left[  1+i\omega_{c}%
\tau\right]  }\times\nonumber\\
&&\int_{0}^{2\pi}\frac{d\varphi}{2\pi}e^{-i\varphi}\left[  \frac{2\epsilon}%
{m}\cos\varphi\partial_{\epsilon}-\frac{\sin\varphi}{m}\partial_{\varphi
}\right]  f_{\omega q}(\epsilon,\varphi).
\end{eqnarray}
Assuming energy independence of $\tau$ and density of states,
$\gamma$, we arrive at
\[
\mathcal{J}=\frac{e^{2}\gamma\tau/m}{\left[  1+i\omega_{c}\tau\right]  }%
\sum_{\pm\omega q}E_{-\omega-q}g_{\omega q}^{0},
\]
thus reducing the problem to that of finding the AC density
modulation $n_{\omega q}=\gamma_{F}g_{\omega q}^{0}$ excited by
the SAW. Note that $\left(  g_{\omega q}^{0}\right)
^{\ast}=g_{-\omega-q}^{0}$, while summation over the SAW harmonics
satisfies $\omega=\mathbf{s}\cdot\mathbf{q}$.

Since the structure of $\mathcal{J}$ repeats that of the Drude
conductivity tensor, it is natural to work with the DC
acousto-electric field generated by the SAW in the direction of
its propagation,
\begin{equation}
\mathbf{E_{\mathrm{AE}}}=\mathbf{\hat{q}}\frac{2}{\epsilon_{F}}\Re\left\{
E_{-\omega-q}g_{\omega q}^{0}\right\}  .
\end{equation}
As we show in the following calculation, the quantity of interest,
$E_{\mathrm{AE}}$, depends on the relaxation rate,
${\tau_{\ast}}^{-1}$, determined by the type of electron
scattering which  is quite different for small-angle ($\sigma$)
scattering from the momentum relaxation rate measured in
conductivity.

The dynamical perturbation $g_{\omega q}\left( \varphi\right)$ can
be found by taking Fourier harmonics of Eq.~(\ref{Kinetic-Eq1}) at
the frequency and wave number of the SAW,
\begin{align}
&  \left[  \partial_{\varphi}-\frac{\mathcal{\hat{C}}_{\sigma/\eta}}%
{\omega_{c}}-i\frac{\omega}{\omega_{c}}+iqR_{c}\cos\varphi\right]  f_{\omega
q}(\epsilon,\varphi)=\nonumber\\
&  =-\frac{(\partial_{\epsilon}f_{T})g_{\omega q}^{0}+f_{\omega q}%
^{0}(\epsilon)}{\omega_{c}\tau_{in}}-\frac{eE_{\omega q}}{\omega_{c}%
}{\mathcal{\hat{P}}}\left[  f_{00}\left(  \epsilon,\varphi\right)
+f_{T}\left(  \epsilon\right)  \right]  .\nonumber
\end{align}
Here we neglect the term $f_{00}\propto\left\vert E_{\omega q}\right\vert
^{2}$\thinspace\ since any resulting corrections in $f_{\omega q}%
(\epsilon,\varphi)$ would be non-linear in the SAW\ power.
Assuming a low temperature regime, $kT\ll\omega_{c}p_{F}/q$, we
integrate the above over energy approximating all energy dependent
parameters (such as $R_{c}$) by their respective values at the
Fermi level, $\partial_{\epsilon}f_{T}\approx -\delta\left(
\epsilon-\epsilon_{F}\right)  $, and arrive at
\begin{equation}
\left[  \partial_{\varphi}-\frac{\mathcal{\hat{C}}_{\sigma/\eta}}{\omega_{c}%
}-i\frac{\omega}{\omega_{c}}+iqR_{c}\cos\varphi\right]  g_{\omega q}%
(\varphi)=\frac{ev_{F}E_{\omega q}}{\omega_{c}}\cos\varphi.\label{Eqn-g-wq}%
\end{equation}
In the limit of $qR_{c}\gg1$, the solution to Eq.~(\ref{Eqn-g-wq})
displays a fast-oscillating angular dependence,
$e^{-iqR_{c}\sin\varphi}$, caused by the last term in brackets on
the LHS of Eq.~(\ref{Eqn-g-wq}). To take those fast oscillations
into account we use the method proposed by Rudin, Aleiner and
Glazman\cite{Aleiner1} and write $g_{\omega q}(\varphi)=h_{\omega
q}(\varphi)e^{-iqR_{c}\sin\varphi}$. Using angular Fourier
harmonics of $h_{\omega q}(\varphi)$, this reads
\begin{equation}
g_{\omega q}=\sum_{k=-\infty}^{\infty}h_{\omega q}^{k}J_{k-n}\left(
qR_{c}\right)  .
\end{equation}
Having multiplied Eq.~(\ref{Eqn-g-wq}) by
$e^{iqR_{c}\sin\varphi-ik\varphi}$ and integrated over angle
$\varphi$, we arrived at the system of coupled equations for
Fourier coefficients $h_{\omega q}^{k}$,
\begin{align}
& \left(  ik-i\frac{\omega}{\omega_{c}}\right)  h_{\omega q}^{k}-\frac
{1}{\omega_{c}}\left\langle e^{iqR_{c}\sin\varphi-ik\varphi}\mathcal{\hat
{C}_{\sigma/\eta}}g_{\omega q}\left(  \varphi\right)  \right\rangle
\nonumber\\
& =\frac{ekE_{\omega q}}{q}J_{k}\left(  qR_{c}\right)  ,\label{h_k-indet}%
\end{align}
where $\langle\ldots\rangle=\int_{0}^{2\pi}d\varphi/2\pi$ stands
for averaging over the angle $\varphi$.

The following analysis of Eq.~(\ref{h_k-indet}) depends on the
form of the collision integral. For isotropic scattering,
\begin{equation}
\left\langle e^{iqR_{c}\sin\varphi-ik\varphi}\mathcal{\hat{C}_{\eta}}g_{\omega
q}\left(  \varphi\right)  \right\rangle =\frac{J_{k}\left(  qR_{c}\right)
g_{\omega q}^{0}}{\tau}-\frac{h_{\omega q}^{k}}{\tau},
\end{equation}
and the elements $h_{\omega q}^{k}$ in equation
Eq.~(\ref{h_k-indet}) decouple. One therefore finds the $m$-th
angular harmonic of $g_{\omega q}(\varphi)$ as
\begin{equation*}
g_{\omega q}^{m}=\sum_{k=-\infty}^{\infty}\frac{J_{k}\left(  qR_{c}\right)
J_{k-m}\left(  qR_{c}\right)  }{ik-i\frac{\omega}{\omega_{c}}+\frac{1}%
{\omega_{c}\tau}}\left\{  \frac{g_{\omega q}^{0}}{\omega_{c}\tau}%
+\frac{ekE_{\omega q}}{q}\right\}  .\label{g^m-isotropic-intermediate}%
\end{equation*}
Setting $m=0$, we find%
\begin{align}
&  g_{\omega q}^{0}=\frac{eE_{\omega q}}{q}\sum_{k=-\infty}^{\infty}%
\frac{kJ_{k}^{2}\left(  qR_{c}\right)  }{ik-i\frac{\omega}{\omega_{c}}%
+\frac{1}{\omega_{c}\tau}}\,\frac{1}{\left[  1-K\right]  }%
,\label{g^0-isotropic-1}\\
&  K=\frac{1}{\omega_{c}\tau}\sum_{k=-\infty}^{\infty}\frac{J_{k}^{2}\left(
qR_{c}\right)  }{ik-i\frac{\omega}{\omega_{c}}+\frac{1}{\omega_{c}\tau}%
}.\label{R-eta}%
\end{align}
In the limit of $qR_{c}\gg1$, then $K\ll1$ since $J_{k}^{2}\left(  qR_{c}%
\gg1\right)  \sim1/qR_{c}\rightarrow0$. Additionally, the linear
$k$ dependence in Eq.~(\ref{g^0-isotropic-1}) may be manipulated
to read $k=-i\left(  ik-i\omega/\omega_{c}+1/\omega_{c}\tau\right)
+i\left( -i\omega/\omega_{c}+1/\omega_{c}\tau\right)  $ - the
first term exactly cancels the resonance denominator, and
application of the identity $\sum _{k}J_{k}^{2}(x)=1$ yields an
approximate form of $g_{\omega q}^{0}$,
\begin{equation}
g_{\omega q}^{0}\approx\frac{eE_{\omega q}}{q}\left\{  \frac{1}{i}%
+\frac{\omega}{\omega_{c}}\sum_{k=-\infty}^{\infty}\frac{J_{k}^{2}\left(
qR_{c}\right)  }{ik-i\frac{\omega}{\omega_{c}}+\frac{1}{\omega_{c}\tau}%
}\right\}  .\label{g0-iso}%
\end{equation}

For small-angle scattering the angle-average in
Eq.~(\ref{h_k-indet}) takes the form \cite{footnote1,Aleiner1}
\begin{align}
&  \left\langle e^{iqR_{c}\sin\varphi-ik\varphi}\mathcal{\hat{C}_{\sigma}%
}g_{\omega q}\left(  \varphi\right)  \right\rangle =-\frac{\left(
qR_{c}\right)  ^{2}}{2\tau}h_{\omega q}^{k}-\frac{\mathcal{G}_{k}}{\tau
},\label{average-temp}\\
&  \mathcal{G}_{k,n}=\left(  \frac{qR_{c}}{2}\right)  ^{2}\left[  h_{\omega
q}^{k-2}+h_{\omega q}^{k+2}\right]  +k^{2}h_{\omega q}^{k}\nonumber\\
&  -\frac{qR_{c}}{2}\left[  \left(  2k+1\right)  h_{\omega q}^{k+1}+\left(
2k-1\right)  h_{\omega q}^{k-1}\right]  .\nonumber
\end{align}
Coupling between different elements $h_{\omega q}^{k}$ in a
combination of Eq.~(\ref{h_k-indet}) and Eq.~(\ref{average-temp})
occurs with multipliers of $\left( qR_{c}\right)
^{2}/\tau\lesssim\omega_{c}$ and $kqR_{c}/\tau\lesssim\omega_{c}$
, and is now much weaker than the coupling one would obtain from a
direct Fourier transform of Eq.~(\ref{Eqn-g-wq}). This enables us
to solve Eq.~(\ref{h_k-indet}) perturbatively in
$\mathcal{G}_{k}$. Note that we also attribute the term
$k^{2}h_{\omega q}^{k}$ to the perturbative correction
$\mathcal{G}_{k}$, since $k\lesssim qR_{c}$, and inclusion of this
term in the leading approximation would exceed the chosen
accuracy. Thus, we write
\begin{align}
&  g_{\omega q}^{m}=\frac{eE_{\omega q}}{q}\sum_{k=-\infty}^{\infty}%
\frac{kJ_{k}\left(  qR_{c}\right)  J_{k-m}\left(  qR_{c}\right)  }%
{ik-i\frac{\omega}{\omega_{c}}+\frac{\left(  qR_{c}\right)  ^{2}}{2\omega
_{c}\tau}}\nonumber\\
&  -\sum_{k=-\infty}^{\infty}\frac{J_{k-m}\left(  qR_{c}\right)  }%
{ik-i\frac{\omega}{\omega_{c}}+\frac{\left(  qR_{c}\right)  ^{2}}{2\omega
_{c}\tau}}\frac{\mathcal{G}_{k}}{\omega_{c}\tau}.\label{g^m-small-angle-long}%
\end{align}
Setting $m=0$, and solving up to second order
Eqs.~(\ref{average-temp}, \ref{g^m-small-angle-long}) in
$\mathcal{G}_{k}$ we find
that in the leading order of the parameters $\omega/qR_{c}\omega_{c}%
=s/v_{F}\ll1$, $k/qR_{c}\lesssim1$ ($qR_{c}\gg1$ and $qR_{c}\ll\omega_{c}\tau
$), the main contribution to the zeroth harmonic $g_{\omega q}^{0}$ is given
by
\begin{equation}
g_{\omega q}^{0}\approx\frac{eE_{\omega q}}{q}\left\{  \frac{1}{i}%
+\frac{\omega}{\omega_{c}}\sum_{k=-\infty}^{\infty}\frac{J_{k}^{2}\left(
qR_{c}\right)  }{ik-i\frac{\omega}{\omega_{c}}+\frac{\left(  qR_{c}\right)
^{2}}{2\omega_{c}\tau}}\right\}\,.  \label{tau-q}%
\end{equation}

Using the similarity of Eqs.~(\ref{g0-iso}, \ref{tau-q}), we
express the SAW
induced electric field, $\mathbf{E}_{\mathrm{AE}}=\mathbf{\hat{q}%
}E_{\mathrm{AE}}$, for both isotropic and small-angle scattering
cases in terms of the effective scattering rate
$\tau^{-1}_{\ast}$,
\begin{eqnarray}
E_{\mathrm{AE}}=\frac{2es\left\vert E_{\omega q}\right\vert ^{2}}{\epsilon
_{F}}\sum_{k=-\infty}^{\infty}\frac{\tau_{\ast}J_{k}^{2}\left(  qR_{c}\right)
}{{\tau_{\ast}}^{2}\left(\omega-k\omega_{c}\right)^{2}+1},\label{E_DC2}%
\\
\tau_{\ast}^{-1}=%
\begin{cases}
\tau^{-1}, & \text{for isotropic scattering},\\
\frac{(qR_{c})^{2}}{2}\tau^{-1}, & \text{for small-angle
scattering\cite{Mirlin1}}.
\end{cases}
\label{tau-star}%
\end{eqnarray}
Geometrical commensurability manifests itself in Eq.~(\ref{E_DC2})
through the appearance of the Bessel function $J_{k}(qR_{c})$.
Additionally, the presence of a finite wave vector lifts selection
rules for electron transitions, allowing transitions otherwise
forbidden by Kohn's theorem\cite{Kohn}, such as resonances at
multiples of the cyclotron frequency.

The dynamical redistribution of electrons leads to screening of
the external SAW field $E_{\omega q}^{\mathrm{SAW}}$ by the 2DEG.
We relate $E_{\omega q}$ to the unscreened field by inclusion of
the dielectric function, $E_{\omega q}=E_{\omega
q}^{\mathrm{SAW}}/\kappa(\omega,q)$. In the Thomas-Fermi
approximation, and in the limit of $qR_{c}>1$ we find that $\kappa
^{-1}(\omega,q)\approx a_{\mathrm{scr}}q$,\cite{Robinson} where
$a_{\mathrm{scr}}=\chi/2\pi e^{2}\gamma$ is the donor-related Bohr
radius ($\chi$ is the background dielectric constant) and
introduce the dimensionless parameter which is a measure of the
amplitude of the screened SAW field normalized by the Fermi
energy,
\begin{equation}
\mathcal{E}=\left(ea_{\mathrm{scr}}E_{\omega q}^{\mathrm{SAW}}
/\epsilon_{F}\right)^{2}\,.
\end{equation}

In the limit of $\omega_{c}\tau_{\ast}\gg1$, we discuss two
extreme cases, $v_{F}\gg s$ and $v_{F}\lesssim s$, where $s$ is
the
SAW\ speed\cite{Willett} in GaAs, $s=2.8\times10^{5}%
\operatorname{cm}%
\operatorname{s}%
^{-1}$. At electron densities $n_e\sim 10^{10}\div 10^{12}%
\operatorname{cm}%
^{-2}$, $v_{F}\gg s$ and the relevant frequency regime in
structures with realistic mobility appears to be $\omega/\omega_c
= (s/v_F)qR_c \ll 1$. In this situation, the largest contribution
to the DC field then comes from the term in Eq.~(\ref{E_DC2}) with
$k=0$, and
\begin{equation}
E_{\mathrm{AE}}\approx\frac{2\mathcal{E}\epsilon_{F}}{es\tau_{\ast}}\frac{J_{0}%
^{2}\left(qR_{c}\right)}{1+(\omega\tau_{\ast})^{-2}}
\label{low-freq-lim}.
\end{equation}

It is interesting to note that the magnetic field dependence of
the amplitude of geometrical oscillations described by
Eqs.~(\ref{low-freq-lim}) strongly differs for the two limiting
types of scattering considered above. In the case of isotropic
(short-range) scatterers the oscillation amplitude decreases as
$N^{-1}$ with the oscillation number $N \sim qR_c/\pi$. In
contrast, for low-angle scattering it is non-monotonic. It
increases linearly in $N$ up to $N^{\sigma}\sim
\sqrt{2\omega\tau}/\pi$ where the oscillations amplitude has a
maximum followed by a gradual $N^{-3}$ decrease.

The result in Eq.~(\ref{E_DC2}) also shows that in a low-density
2DEG or in a gas of ``heavy'' composite fermions\cite{Willett},
such that $v_{F}\lesssim s$, resonances in the AE\ effect at
 $\omega=k\omega_c$ become possible. Due to the oscillatory behaviour
 of Bessel functions $J_{k}(qR_c)$ at $qR_{c}\gg 1$ and since
 $\frac{\omega}{\omega_c}= \frac{s}{v_F}qR_c$, resonances
 would appear in the experiment as a sequence of Lorentzians
 of apparently random height.

To study the damping of geometrical oscillations at low magnetic
fields $\omega_c\tau_{\ast}<1$, we use the method of residues,
transforming the summation in Eq.~(\ref{E_DC2}) into the integral
\[
E_{\mathrm{AE}}\approx\frac{\mathcal{E}\omega p_{F}/2\pi i}{e\omega_{c}%
\tau_{\ast}}\oint_{C}\frac{\left[  1+\sin\left(  2qR_{c}-z\pi\right)  \right]
\cot(\pi z)dz}{[z-\frac{\omega}{\omega_{c}}+\frac{i}{\omega_{c}\tau_{\ast}%
}][z-\frac{\omega}{\omega_{c}}-\frac{i}{\omega_{c}\tau_{\ast}}]}%
\]
where for $qR_{c}\gg1$, $J_{k}\left(  qR_{c}\right)
\approx\sqrt{2/\pi qR_{c}}\cos(qR_{c}-k\pi/2-\pi/4)$, and the
contour $C$ consists of two parts: in the upper half-plane,
$C_{+}=x+i0$ and in the lower half-plane, $C_{-}=x-i0$. Each
contour, $C_{\pm}$ is then moved away from the real axis,
$C_{\pm}\rightarrow C_{\pm}^{\prime}=x\pm i|y|$, such that
$e^{-2|y|}\ll1$, when each contour picks up exactly one residue
from the poles at $z_{\pm}=\omega/\omega_{c}\pm i/\omega_{c}\tau
_{\ast}$ (here and below the subscript $\pm$ is determined by the
subscript of the contour). The numerator of the integrands in the
shifted line integrals $\int_{C_{\pm}}dz$ are approximated using
$e^{-2|y|}\ll1$, thus yielding $\cot(\pi z)\approx\mp i$ and
$\sin\left(  qR_{c}-z\right)  \approx\pm e^{\pm(2iqR_{c}-iz)}/2i$.
After this, each contour is then moved, $C_{\pm
}^{\prime}\rightarrow C_{\pm}^{\prime\prime}$, such that $\Im C_{\pm}%
^{\prime\prime}\rightarrow\mp\infty$, passing the real axis as
they approach the opposite extremes of the complex plane. Thus, we
arrive at
\[
E_{\mathrm{AE}}=\frac{\mathcal{E}\omega p_{F}}{e}\left\{  1+\sin\left(
2qR_{c}-\frac{\pi\omega}{\omega_{c}}\right)  e^{-{\pi}/{\omega_{c}\tau_{\ast}%
}}\right\}  .
\]

The latter equation is typical for damped geometrical
oscillations\cite{Mirlin1}. It shows how commensurability
oscillations die away when $\pi/\omega_{c}\tau_{\ast}\gtrsim1$ and
that the onset of oscillations occurs at the magnetic field value
such that
\begin{equation}
R_c<R_c^{\ast}\sim\left\{
\begin{array}
[c]{ll}%
{l/\pi}& \quad \text{for isotropic scattering,}%
\\
{l\sqrt[3]{2/\pi(lq)^2}}&\quad \text{for small-angle scattering.}
\end{array}
\right.  \nonumber\label{B-crit}%
\end{equation}
Together with the result in Eq.~(\ref{low-freq-lim}), the latter
offset condition shows that the number of $B^{-1}$ oscillations,
$N\sim qR_c^{\ast}/\pi$ detectable in a sample with the mean free
path $l=v_F\tau \gg 2\pi/q$ is larger when its mobility is limited
by short-range scatterers ($N<N^{\eta}$) than when scattering is
due to smooth disorder ($N<N^{\sigma}$), where
\begin{equation}
N^{\eta} \sim \frac{ql}{\pi}\,,\quad \text{vs} \quad
N^{\sigma}\sim
\mathrm{min}\left\{(ql)^{\frac{1}{3}}\,,\sqrt{\frac{sql}{\pi
v_F}}\right\}.\label{Nmax}
\end{equation}

In the regime of damped oscillations a finite and apparently field
independent AE\ effect persists up to the field $v_{F}B>E_{\omega
q}$ (when channelling takes it toll),
\begin{equation}
E_{\mathrm{AE}}\approx e\omega p_{F}\left(a_{\mathrm{scr}}E_{\omega q}^{\mathrm{SAW}}%
/\epsilon_{F}\right)^{2}.\label{final}%
\end{equation}

The above presented analysis explains why the observed magnetic
field dependence of the AE effect by Shilton \textit{et
al.}\cite{Shilton} contained only one pronounced oscillatory
feature associated with geometrical commensurability before the AE
effect saturated at a finite value in low magnetic fields. This
contrasted a comparison that the author of [\cite{Shilton}] made
with the SAW absorption modelled in the $\tau$-approximation which
suggested that many ($N\sim 3\div 10$) oscillations should become
visible while varying the SAW wavelength from $\lambda \sim 10
\operatorname{\mu m}$ to $\lambda \sim 3 \operatorname{\mu m}$
against a mean free path of $l=30\operatorname{\mu m}$. In
structures with low-angle scattering $N^{\sigma}$, in
Eq.~(\ref{Nmax}) increases very slowly with the increase of SAW
wave number remaining $N \sim 1 \div 2$ for all three SAW sources
used in Ref. [\cite{Shilton}]. In the other AE experiment with
surface acoustic waves\cite{Esslingen} that we are aware of, the
mobility of samples was not sufficient to observe geometrical
oscillations, though the saturated non-oscillatory low-field AE
was seen.

In conclusion, we present a microscopic theory of the
acousto-electric effect generated by surface acoustic waves in a
high mobility 2D electron gas with either isotropic or small-angle
impurity scattering. In both cases the acousto-electric effect
exhibits Weiss oscillations periodic in $B^{-1}$ due to the
commensurability of the SAW period with the size of the cyclotron
orbit and resonances as a function of SAW frequency at multiples
of the cyclotron frequency. We find that the geometrical and
frequency resonances vanish in low fields where
$\omega_{c}\tau_{\ast}\lesssim 1$ (with the time scale
$\tau_{\ast}$ dependent on the type of scattering), however its
non-oscillatory part is finite to the lowest fields.

We thank\ I. Aleiner, J. Bergli, Y. Galperin, M. Kennett and A.\
Mirlin for discussions and acknowledge support from EPSRC grant
GR/R17190, Lancaster Portfolio Partnership and INTAS Network
0351-6453.


\begin{thebibliography}{99}
\bibitem {footnote1}The average is calculated by turning the operator
$\mathcal{\hat{C}_{\sigma}}\propto\partial_{\varphi}^{2}$ onto the term
$e^{iqR_{c}\sin\varphi-ik\varphi}$ which produces a constant and sinusoidal
coefficients of single and double frequency. Turning these coefficients into
complex exponentials and using the relation $J_{k-n}(qR_{c})=\int_{0}^{2\pi
}d\varphi e^{iqR_{c}\sin\varphi-i(k-n)\varphi}/2\pi$ gives the matrix elements
$h_{\omega q}^{k}$, $h_{\omega q}^{k\pm1}$ and $h_{\omega q}^{k\pm2}$ respectively.
\end{thebibliography}
\end{document}